# Mueller matrix spectroscopy of Fano resonance in Plasmonic Oligomers

Shubham Chandel[1], Ankit K Singh[1], Aman Agrawal[2], Aneeth K A[1], Angad Gupta[1], Achanta Venugopal[2]* and Nirmalya Ghosh[1]*

[1] *Department of Physical Sciences, Indian Institute of Science Education and Research Kolkata, West Bengal, 741246, India*

[2] *Tata Institute of Fundamental Research, Mumbai 400005, INDIA*

Fano resonance in plasmonic oligomers originating from the interference of a spectrally broad superradiant mode and a discrete subradiant mode is under intensive recent investigations due to numerous potential applications. In this regard, development of experimental means to understand and control the complex Fano interference process and to modulate the resulting asymetric Fano spectral line shape is highly sought after. Here we present a polarization Mueller matrix measurement and inverse analysis approach for quantitative understanding and interpretation of the complex interference process that lead to Fano resonance in symmetry broken plasmonic oligomers. The spectral Mueller matrices of the plasmonic oligomers were recorded using a custom designed dark-field Mueller matrix spectroscopy system. These were subsequently analyzed using differential Mueller matrix decomposition technique to yield the quantitative sample polarimetry characteristics, namely, polarization diattenuation (d) and linear retardance ($\delta$) parameters. The unique signature of the interference of the superradiant dipolar plasmon mode and the subradiant quadrupolar mode of the symmetry broken plasmonic oligomers manifested as rapid spectral variation of the diattenuation and the linear retardance parameters across the Fano spectral dip. The polarization information contained in the Mueller matrix was further utilized to desirably control the Fano spectral line shape. The experimental Mueller matrix analysis was complemented with finite element based numerical simulations, which enabled quantitative understanding of the interference of the superradiant and the subradiant plasmon modes and its link with the polarization diattenuation and retardance parameters.

The asymmetric Fano resonance originating from the interference of a narrow resonance with a broad spectra or continuum of states is a universal phenomenon, observed in diverse spectroscopic systems[1-4]. Studies of Fano resonance in plasmonic and other nano optical systems are crucial in the context of fundamental understanding of the interference effect that leads to Fano resonance[4-10] and for plethora of potential applications like sensing, switching, invisibility cloaking, slow light devices, lasing and many more[11-17]. Among the various plasmonic nanostructures, Fano resonance in plasmonic oligomers has been extensively studied because of the rich underlying physics and its potential as a sensing device[18-20]. Fano resonance in oligomers appears due to the interference of subradiant and superradiant plasmon modes[10]. The superradiant mode originates due to the sum of dipole moments of the individual nanoparticles in the oligomer, whereas the subradiant mode originates from the cancelation of net dipole moments of the individual nanoparticles[10]. The formation and the nature of the superradiant and subradiant modes in plasmonic oligomers crucially depend on the spatial distribution of the electric field of light, which can be controlled by the polarization state. While the symmetric oligomers have polarization insensitive response, the symmetry broken oligomers may exhibit wealth of polarization characteristics. Indeed, initial studies with selected linear polarization detection have shown interesting anisotropic response in symmetry broken plasmonic oligomers [18]. However, a quantitative study of the polarization response is not yet available. Such a study would help in the optimization of polarization states for tunable Fano resonance in a single device for various applications.

We have therefore studied complete polarization response of plasmonic oligomers using a homebuilt polarimetry platform based on a Dark Field microscope, which is capable of recording full scattering Mueller matrix over broad spectral range[21]. Mueller matrix is a 4x4 matrix which incorporates all the polarization properties of the sample in its various elements[21,22]. The individual sample polarization properties encoded in Mueller matrix can be extracted and quantified in terms of the fundamental polarization parameters, namely, diattenuation, retardance and depolarization. Diattenuation is the differential attenuation of orthogonal polarizations (reflects amplitude anisotropy effect), whereas retardance stands for phase difference between two orthogonal polarizations (incorporates phase anisotropy) and the depolarization is defined as loss of degree polarization[8,21,22]. Extraction and quantification of these polarimetry effects can be achieved by polar decomposition of Mueller matrix (and its other variants) or by differential Mueller matrix decomposition analysis[23,24]. The latter technique (used here) has an advantage that it can tackle complex systems exhibiting simultaneous occurrences of many polarization effects that too in presence of scattering. We have chosen symmetry broken oligomers (heptamer) to study the polarization effects and their relations with the Fano interference. The results yielded interesting link between the Mueller matrix-derived polarization parameters and the spectral interference of the subradiant and the superradiant plasmon modes. It is further demonstrated that the polarization information stored in the Mueller matrix can be used to desirably tune the Fano spectral lineshape. Here, we are presenting two interesting aspects of polarimetric investigation of Fano resonance in plasmonic oligomers – (a) how the Mueller matrix-derived polarimetry parameters can be used to gain quantitative understanding on the Fano interference process, and (b) how the information encoded in the Mueller matrix can be utilized to control and tune Fano resonance via polarization. To the best of our knowledge this is the first experimental demonstration of full spectral Mueller matrix measurements from plasmonic oligomers and its inverse analysis using differential Mueller matrix formalism.

The core of our experimental system is a custom designed setup integrating a dark field microscope that employs excitation with broadband light and subsequent recording of sixteen polarization resolved scattering spectra (wavelength $\lambda = 400 - 725$ nm with 1 nm resolution) by sequential generation and analysis of four optimized elliptical polarization states (see Supporting information). The experimental system enables recording of the 4×4 spectral Mueller matrix $\mathbf{M}(\lambda)$ with high precision (enabled by a robust Eigen value calibration method) [21]. We fabricated plasmonic oligomer heptamer, consisting of six nano-disks making a ring around a central nano-disk. The symmetry broken oligomer was fabricated by taking one of the ring particles away leaving the system asymmetric along one of the orthogonal axis. The fabrication involved Electron beam lithography and metal deposition by thermal evaporation technique (see Supporting information). The geometrical parameters of the oligomer were - inner disk diameter 100nm, outer disk diameter 83nm, disk separation 32nm. The heptamer array had a period of 900nm. The recorded scattering Mueller matrix $M(\lambda)$ is decomposed using differential matrix decomposition which assumes simultaneous occurrence of individual polarization and depolarization effects[23,24]. Using this approach, the elementary polarization properties are extracted by constructing the Lorentz antisymmetric (Lm), and symmetric (Lu) matrices from the logarithm of the Mueller matrix $M$ (see Supporting information). The mean values of the elementary polarization properties are represented by the off-diagonal elements of Lm matrix, while those of Lu matrix express respective uncertainties/standard deviations of the polarization parameters. Additionally, the diagonal elements of the matrix $L_u$ represent the depolarization coefficients. The details of the decomposition process are given in the supporting information and elsewhere[23,24].

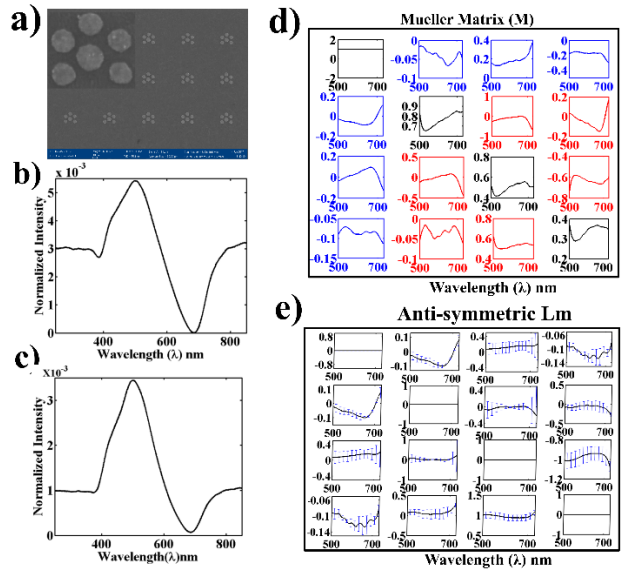

Figure 1: (a) SEM image of typical symmetry broken plasmonic oligomer heptamer, (b, c) polarization blind scattering spectra from the symmetric and symmetry broken oligomers. (d) The spectral scattering Mueller matrix $M$ ($\lambda$) (normalized by M11($\lambda$ )) of symmetry broken plasmonic oligomer. The constituent polarimetric properties are encoded as: diattenuation in the first row and column (blue); retardance in the off-diagonal elements of the lower 3 × 3 block (red), depolarization as diagonal elements (black). (e) The wavelength dependence of the decomposed anti-symmetric Lm matrix. The standard deviations of the off-diagonal elements are represented as error bars from the corresponding Lu matrix.

Clear Fano resonance can be seen in both the spectra of symmetric and symmetry broken oligomers with a prominent dip around $\lambda_{Fano}$ ~ 690nm (Fig. 1b and 1c). As previously shown, the Fano resonance in such oligomer structure originates due to the interference of a broad dipolar superradiant mode and a dark subradiant mode[18]. In case of the symmetric oligomer, the subradiant mode is primarily dipolar in nature. In contrast, for the symmetry broken oligomers, the subradiant mode exhibits admixture of nanoparticle quadrupoles due to the reduced symmetry. Moreover, the contribution of the quadrupoles in the subradiant mode also critically depend upon the polarization state because the direction of the polarization vector with respect to the symmetry axis of the oligomer decides the dipolar and quadrupolar contribution of the subradiant mode[18]. The non-zero magnitudes of almost all the off-diagonal Mueller matrix elements of the symmetry broken oligomer (Figure 1d) underscore the anisotropic nature of Fano resonance. Specifically, the $M_{34}/M_{43}$, $M_{24}/M_{42}$ elements and $M_{12}/M_{21}$, $M_{13}/M_{31}$, $M_{14}/M_{41}$ elements characteristically reflect the phase and the amplitude anisotropy effects of Fano resonance. Interestingly, these anisotropy carrying elements exhibit sharp spectral variations around the Fano spectral dip (690 nm). Additionally, the diagonal elements have values less than unity signifying the presence of depolarization due to incoherent averaging of polarization signals over many scattering angles in our high NA microscopic geometry. This depolarization effect is subsequently taken care of by the differential Mueller matrix decomposition which isolates the anisotropy parameters (via the Lm matrix) from the depolarization (diagonal elements of the Lu matrix). The resulting elements of the matrix Lm (Fig. 1e) exhibit accumulated amplitude anisotropy (diattenuation encoded in Lm(1,2), Lm(1,3) and Lm(1,4)) and phase anisotropy effects (linear retardance encoded in Lm(2,4), Lm(3,4)). The standard deviations in the anisotropy parameters (expressed in off-diagonal elements of Lu) arise due to orientation averaging affects in our high NA experimental geometry.

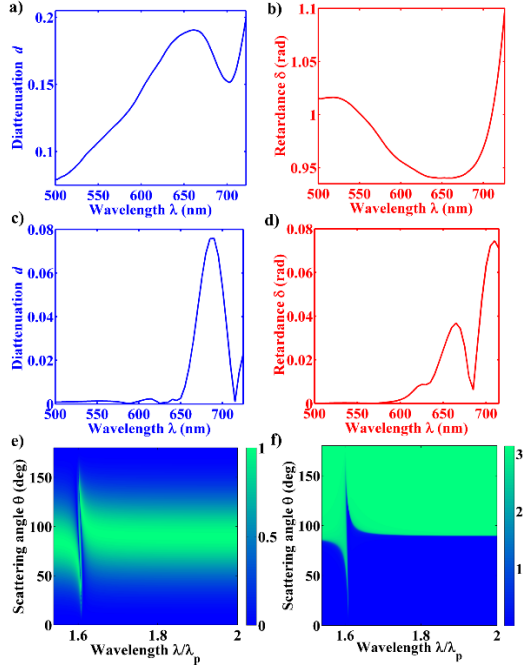

Figure 2: Wavelength dependence of the polarization parameters, (a) diattenuation and (b) linear retardance obtained via differential matrix decomposition of experimental Mueller matrix from symmetry broken plasmonic oligomers (corresponding to Fig. 1d and 1e). (c, d) The corresponding spectral variations of the polarization parameters obtained from finite element method based simulations. (e, f) The normalized wavelength ($\lambda/\lambda_p$) variation of the diattenuation and linear retardance parameters for a spherical nanoparticle (diameter = 33.422 nm) of an ideal Drude metal exhibiting dipolar and quadrupolar plasmon resonances.

The magnitudes of diattenuation (d) (Fig. 2a) and linear retardance (δ) (Fig. 2b) extracted from the Lm matrix exhibit rapid variation across the Fano spectral dip ($\lambda_{Fano}$ ~ 690 nm) indicating the strong anisotropic nature of Fano resonance. Numerical simulations of the scattering Mueller matrices of the symmetry broken plasmonic oligomer were performed using finite element method (FEM) with COMSOL software. The d and the δ parameters extracted from the simulated Mueller matrices (Fig. 2c and 2d) exhibit qualitatively similar sharp variations across the spectral dip of Fano resonance. Differences in the experimental and simulated parameters are due to the fact that the simulation was done for plane wave excitation whereas the experimental system used tightly focused light beam, which resulted in additional effects on averaging of polarized intensities over finite scattering angles. Since the interference of the dipolar (superradiant) and the quadrupolar (subradiant) modes is crucial towards the Fano resonance and its polarization response in symmetry broken oligomers, it might be useful to gain some insight on the effect of the dipolar and the quadrupolar modes on the polarization parameters for a simpler system. Thus, we have considered interfer ence of a dipolar and a quadrupolar plasmon mode in scattering from a spherical nanoparticle (diameter = 33.422 nm) of an ideal Drude metal (dielectric permittivityϵ is described by the Drude formula, γ≈0.0239, where γ is the collision frequency and $\lambda_p$=150 nm is the plasma wavelength). Figure 2e and 2f present the results for the d and δ parameters derived from the scattering matrix generated using Mie theory. The results show that both the d and δ parameters exhibit sharp variation across the wavelength of the quadrupolar resonance ($\frac{\lambda}{\lambda_p} = 1.604$), where the spectrally broad dipolar plasmon mode overlaps and interfere with the spectrally narrower quadrupolar plasmon mode. The slowly (with wavelength) varying phase of the broad dipolar mode and the steeply changing phase of the quadrupolar mode lead to sharp phase difference between the orthogonal linear polarizations in the overlap spectral region, manifesting as rapidly varying spectral linear retardance effect. Similarly, differential excitation of the dipolar and the quadrupolar modes by orthogonal linear polarizations also leads to sharp spectral variation of diattenuation across the peak of the quadrupolar resonance. The observed rapid spectral variation of d and δ parameters across the Fano spectral dip of the symmetry broken plasmonic oligomer thus appears to be a characteristic signature of the interference of a dipolar and a quadrupolar mode. Thus, the Mueller matrix-derived diattenuation (d) and linear retardance (δ) parameters capture and quantify fundamentally interesting information on the interference of the superradiant dipolar mode and the subradiant quadrupolar mode of symmetry broken plasmonic oligomers and on the resulting anisotropic nature of Fano resonance in such systems.

The d(λ) and δ(λ) parameters may also serve as efficient experimental metric for sensing applications. This follows because these parameters show sharp spectral features that are extremely sensitive to the local dielectric environment. This is demonstrated in Figure 3a, where the spectral variations of d(λ) and δ(λ) show dramatic changes and extreme sensitivity (*Δλ/Δn* ~250 nm/RIU) towards a change in the refractive index of the surrounding medium. These results are generated using FEM simulations of the symmetry broken plasmonic oligomers (same geometrical parameters as in Fig.2). These parameters therefore hold considerable promise as fundamentally interesting, experimentally accessible parameters for studying anisotropic Fano resonances in diverse plasmonic systems and sensing applications exploring these novel experimental polarization metrics may also be envisaged.

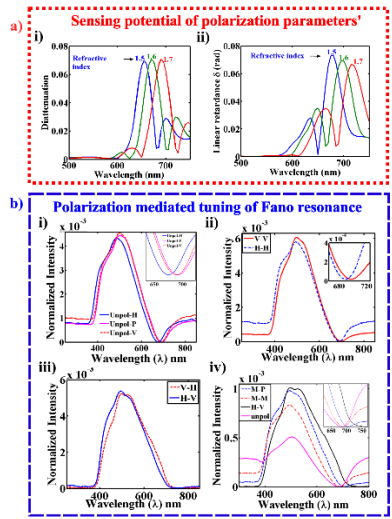

Figure 3: **(a): Potential applications of polarization diattenuation and linear retardance parameters as sensing metrics.** (a)The wavelength variation of theoretically computed d(λ) **(i)** and δ(λ) **(ii)** parameters for a symmetry broken plasmonic oligomer sample with varying refractive index, generated from FEM based simulations. **(b)**: **Polarization controlled tuning of Fano resonance using Mueller matrix spectroscopy**. **i)** Spectral variation of polarized detection, inset shows the zoomed in region of the long wavelength dip position. **ii)** Polarized pre-post selection with co-polarization of Horizontal and vertical, and **iii)** with cross polarization presenting the presence and absence of Fano interference/ spectral profile. **iv)** Polarized pre-post selection with optimized polarization states giving rise to tuning of Fano spectral dip.

We now turn to how the information encoded in the Mueller matrix can be utilized to control and tune Fano resonance. For this purpose, we pre and post select polarization states on the experimental Mueller matrix $M(\lambda)$ to generate the resulting spectral intensity profile. as $I_s(\lambda) = \frac{1}{2} \boldsymbol{S}_{out}^T M(\lambda) \boldsymbol{S}_{in}$. Here, $\boldsymbol{S}_{in}$ and $\boldsymbol{S}_{out}$ are the 1x4 real Stokes vectors corresponding to the input and the projected polarization states of light. For excitation with horizontally polarized light and subsequent detection of horizontally polarized fraction of the scattered light ($\boldsymbol{S}_{in} = \boldsymbol{S}_{out} = [1, 1, 0, 0]^T$), the Fano spectral dip is observed to shift towards shorter wavelength ($\lambda_{Fano} \sim 685\ nm$) as compared to that observed for excitation and detection with vertical polarization ($\boldsymbol{S}_{in} = \boldsymbol{S}_{out} = [1, -1, 0, 0]^T$) ($\lambda_{Fano} \sim 700\ nm$) (Fig. 3 b(ii)). This can be explained by noting that the symmetry axis of the plasmonic oligomer lies along the vertical polarization direction, in which case, both the superradiant and the subradiant modes are primarily dipolar in nature[18]. In contrast, for excitation with horizontal polarization, the contribution of quadrupolar mode arises due to the lack of the symmetry axis along the polarization vector. The contribution of the quadrupolar subradiant mode leads to a shift of the Fano spectral dip towards shorter wavelength. The spectral position of the peak on the other hand remained relatively unaltered because the superradiant dipolar mode is only weakly influenced by the polarization state. Note that this shift in the Fano spectral dip between horizontal and vertical polarization excitation is also responsible for the previously observed sharp variation of the polarization diattenuation parameter across the Fano spectral dip. Pre and post selection with mutually orthogonal polarizations (e.g., horizontal-vertical or vice versa) destroys the interference of the resonant modes, leading to disappearance of the Fano spectral dip (Fig. 3b(iii)). Thus, with appropriate pre and post selection of the polarization states, one may control the hybridization of plasmon modes and modulate the resulting interference of the modes in symmetry broken oligomers leading to tuning of the Fano spectral dip, as illustrated further in Fig. 3 b(iv). A large tunability of $\lambda_{Fano}$ (~ 90 nm) is demonstrated with other pre and post selection of linear polarizations (+45 degree, $\boldsymbol{S} = [1, 0, 1, 0]^T$ denoted as P; and -45 degree, $\boldsymbol{S} = [1, 0, -1, 0]^T$ denoted as M state). These results demonstrate that state of polarization of light can be used as a unique handle to modulate and control the interference of the contributing modes (via spectral Mueller matrix) in Fano resonance resulting in tunability of Fano spectral line shape.

We have conducted polarization Mueller matrix spectroscopic studies on Fano resonance in symmetry broken plasmonic oligomers. The rich polarization information encoded in the Mueller matrix was gleaned and quantified using differential matrix decomposition. The analysis yielded intriguing polarization diattenuation (d(λ)) and linear retardance (δ(λ)) effects, which exhibited rapid variation across the spectral dip of Fano resonance. Such distinctive spectral features of these polarization parameters and the corresponding anisotropic Fano resonance were linked to differential excitation and differential phase of the dipolar and the quadrupolar superradiant and subradiant (respectively) plasmon modes in symmetry broken oligomers. The d(λ) and δ(λ) parameters thus encoded potentially valuable information on the relative amplitude and phases of the interfering modes in Fano resonance. They also appeared extremely sensitive to the local dielectric environment and hold considerable promise as novel sensing metrics. The polarization information contained in the spectral Mueller matrices was further utilized to modulate the Fano spectral line shape by pre and post selection of appropriate polarization states. The demonstrated tunability and control of the spectral Fano dip in a single device of coupled plasmonic system may boost active Fano resonance-based applications e.g., by developing polarization controlled Fano devices for applications in multi-sensing, filtering and color display, switching and so forth.